\documentclass{iopconfser}

\usepackage{amsmath}
\usepackage{graphicx}
\usepackage[numbers,square,sort&compress]{natbib}
\usepackage{hyperref}
\usepackage{braket}
\usepackage[up,bf,raggedright]{titlesec}
\usepackage{xcolor}

\definecolor{customred}{RGB}{189,37,44}

\titleformat{\section}
  {\normalfont\Large\bfseries\color{customred}}{\thesection.}{0.15em}{\vspace{12pt}}
  
\titleformat{\subsection}
  {\normalfont\bfseries}
  {\thesubsection.}{0.15em}{\vspace{3pt}}

\begin{document}

\title{Challenging Excited States from Adaptive Quantum Eigensolvers: Subspace Expansions vs. State-Averaged Strategies}

\author{Harper R. Grimsley$^{1}$ and Francesco A. Evangelista$^{1}$}

\affil{$^{1}$Chemistry Department, Emory University}

\email{hgrimsl@emory.edu, francesco.evangelista@emory.edu}

\begin{abstract}
	The prediction of electronic structure for strongly correlated molecules represents a promising application for near-term quantum computers. 
	Significant attention has been paid to ground state wavefunctions, but excited states of molecules are relatively unexplored.  
	In this work, we consider the ADAPT-VQE algorithm, a single-reference approach for obtaining ground states, and its state-averaged generalization for computing multiple states at once.  
	We demonstrate for both rectangular and linear H$_4$, as well as for BeH$_2$, that this approach, which we call MORE-ADAPT-VQE, can make better use of small excitation manifolds than an analagous method based on a single-reference ADAPT-VQE calculation, q-sc-EOM.
	In particular, MORE-ADAPT-VQE is able to accurately describe both avoided crossings and crossings between states of different symmetries.
	In addition to more accurate excited state energies, MORE-ADAPT-VQE can recover accurate transition dipole moments in situations where traditional ADAPT-VQE and q-sc-EOM struggle.
	These improvements suggest a promising direction toward the use of quantum computers for difficult excited state problems. 
\end{abstract}


\section{Introduction}\label{introduction}
Excited states of molecules are crucial to understanding chemical phenomena, including Jahn--Teller distortions~\cite{jahn_stability_1937}, molecular spectra~\cite{atkins_molecular_2011}, and photochemical reactions~\cite{turro_modern_1978}.
Photochemical processes often involve conical intersections or avoided crossings. 
Accurately predicting such features is therefore critical in an excited state method.
A common solution is to use the state-averaged~\cite{werner_quadratically_1981} complete active space self-consistent field (CASSCF)~\cite{roos_complete_1980} method to find $k$ high-quality, orthogonal states whose energies are each bound from below by the $k$-lowest exact energy.
This is accomplished by selecting an ($N$, $M$) active space of $N$ electrons and $M$ spatial molecular orbitals.
In complete active space configuration interaction (CASCI), a diagonalization of the Hamiltonian is performed in the basis of excitations within this active space, leaving the rest of the orbitals either always occupied or always unoccupied.
CASSCF removes the ambiguity of active space selection by simultaneously optimizing the CI coefficients within the active space and the CI energy as a function of all the orbitals.
In state-averaged CASSCF, one instead minimizes a weighted average of the $k$ CI solutions within the active space which are lowest in energy.
Unfortunately, CASSCF cannot be feasibly applied to large active spaces, motivating a quantum alternative.

Within the field of quantum computing, various approaches have been proposed for obtaining acceptable approximations to both ground and excited state molecular energies.
These include quantum phase estimation (QPE)~\cite{kitaev_quantum_1995,aspuru-guzik_simulated_2005},
quantum subspace diagonalization methods~\cite{mcclean_hybrid_2017,motta_determining_2020,qfd,stair_multireference_2020,tkachenko_quantum_2024} 
quantum imaginary time evolution (QITE)~\cite{motta_determining_2020},
the quantum alternating operator ansatz (QAOA)~\cite{kremenetski_quantum_2021},
the projective quantum eigensolver (PQE)~\cite{stair_simulating_2021},
quantum approaches based on the contracted Schr{\"o}dinger equation (CSE)~\cite{smart_quantum_2021,benavides-riveros_quantum_2024},
the linear active space unitary coupled cluster (LAS-UCC) method~\cite{otten_localized_2022},
the ``power of sine Hamiltonian'' (PSHO) method~\cite{xie_power_2022},
and the full quantum excited-state solver (FQESS)~\cite{wen_full_2024},
among others.
Of particular relevance to this work is the variational quantum eigensolver (VQE)~\cite{peruzzo_variational_2014, mcclean_theory_2016}, one of the most ubiquitous and intuitive methods for efficiently using quantum resources. 
A difficult task in VQE is to choose an efficient ansatz structure to describe the ground state wavefunction, as will be explained in more detail in section \ref{theory}.  
A historically effective approach to solving this problem, the Adaptive, Problem-Tailored (ADAPT-) VQE framework, iteratively constructs the ansatz from primitive unitaries~\cite{grimsley_adaptive_2019}.

While useful for obtaining ground states, ADAPT-VQE generally requires a complementary, non-iterative algorithm to describe excited states, though other schemes such as variational quantum deflation~\cite{chan_molecular_2021} and $\Delta$ADAPT-VQE~\cite{nykanen_toward_2024} are also possible. 
Various methods can act as complements to ADAPT-VQE, including Quantum Subspace Expansion (QSE)~\cite{mcclean_hybrid_2017}, quantum Equation-Of-Motion (qEOM)~\cite{ollitrault_quantum_2020}, and the quantum, self-consistent Equation-Of-Motion (q-sc-EOM)
\cite{asthana_quantum_2023} methods.
In addition to excited state energies, linear response approaches derived from these methods show promise for the calculation of molecular properties ~\cite{kumar_quantum_2023,reinholdt_subspace_2024,ziems_which_2024,buchwald_reduced_2024}.

A fundamental problem with subspace expansions, however, is that the ground state energy is obtained at a different level of theory from the excited state energies, and it is difficult to ensure that the expansion basis will be appropriate for describing the excited states.
For example, a manifold generated by applying all excitation operators with truncated rank may miss important contributions from higher-body terms and, at the same time, incur a high cost due to the need to evaluate a number of matrix elements that scales quadratically with the number of EOM excitation operators.
Therefore, state-averaged methods that use a few chemically informed reference states are potentially advantageous in terms of reducing quantum resources and sensitivity to noise.
In this work, a Multistate-Objective, Ritz-Eigenspectral (MORE-) ADAPT-VQE, addresses these problems by building its ansatz and optimizing its parameters based on the average energy of multiple states.
The states obtained this way are then used as the basis for a Ritz diagonalization of the Hamiltonian to approximate the individual eigenstates.

While several groups have investigated the idea of using state-averaged VQE for a fixed ansatz structure~\cite{nakanishi_subspace-search_2019,parrish_quantum_2019,yalouz_state-averaged_2021}, only Ref.~\citenum{fitzpatrick_self-consistent_2024} has considered state-averaged construction of the ansatz itself in the ADAPT-VQE framework.
During the preparation of this manuscript, a revised version of Ref.~\citenum{fitzpatrick_self-consistent_2024} was published, and the use of a shared ansatz built by state-averaging ADAPT-VQE energy gradients was explicitly described. Here, we extend this work by comparing state-averaged ADAPT-VQE to a subspace expansion about the ground state from traditional ADAPT-VQE and also examine the quality of the corresponding transition dipole moments.
If one ignores orbital optimization, their method, the state-averaged (SA) ADAPT-VQE-SCF algorithm, becomes equivalent to the strategy considered in this work.
To distinguish these two approaches, we denote the state-averaged version of ADAPT-VQE without orbital optimization as MORE-ADAPT-VQE.
It is worth noting that a recent generalization~\cite{benavides-riveros_quantum_2024} of the quantum ACSE~\cite{smart_quantum_2021} method bears many similarities to this approach, despite not using the ADAPT-VQE framework.  

Here, we concern ourselves primarily with the comparison of MORE-ADAPT-VQE to approaches based on a single-reference ADAPT-VQE ansatz, isolating the effects of parallel ansatz construction from those of state-averaged orbital optimization.
Additionally, we consider much larger sets of excited states and compute the potential energy curves of multiple test systems, as well as the transition dipole moments of BeH$_2$.

The remainder of this work is divided as follows:  
In section \ref{theory}, we outline the theory of VQEs and q-sc-EOM and describe in detail the MORE-ADAPT-VQE method. 
In section \ref{results}, we consider molecules with representative features of photochemistry problems, including both avoided and realized crossings between states.
We report potential energy curves obtained from MORE-ADAPT-VQE, comparing them to those obtained using q-sc-EOM expansions about single-reference ADAPT-VQE calculations.
For the insertion of beryllium into H$_2$, we also compare transition dipole moments obtained by each method.
Finally, in section \ref{conclusions}, we summarize the outlook of the method and outline future research directions.

\section{Theory}\label{theory}
\subsection{Variational Quantum Eigensolvers}
Existing quantum hardware faces various problems, including qubit decoherence, a trend which is expected to remain the case throughout a protracted ``noisy, intermediate-scale, quantum'' (NISQ) era~\cite{preskill_2018_quantum}.
Due to this decoherence, only short circuits can be implemented, making it impossible to realize attractive algorithms like quantum phase estimation for interesting systems.
VQEs were designed to alleviate this problem by introducing a short, parameterized, unitary operator $\hat{{U}}\left(\boldsymbol\theta\right)$\cite{peruzzo_variational_2014,mcclean_theory_2016}.
The parameters are real numbers which can be represented as an $N$-dimensional vector $\boldsymbol{\theta} \in \mathbf{R}^{N}$ and are classically optimized to minimize the energy functional
\begin{align}
	E\left(\boldsymbol\theta\right) = \braket{\phi_0|\hat{U}^\dagger\left(\boldsymbol\theta\right)\hat{H}\hat{U}\left(\boldsymbol\theta\right)|\phi_0},
\end{align}
where $\ket{\phi_0}$ is a predefined reference state.
The actual measurements are performed on a quantum computer by repeatedly realizing $\hat{U}\left(\boldsymbol\theta\right)\ket{\phi_0}$ as a quantum circuit and measuring individual terms of the Hamiltonian in the computational basis. 
The core idea of the method is to outsource as much work as possible to classical computers, using the quantum processor only for state preparation and measurement, which cannot be efficiently simulated on a classical computer. 

The VQE formalism has no constraints on the structure of $\hat{U}$, and there exist numerous approaches with different fixed ansatz designs~\cite{peruzzo_variational_2014,wecker_progress_2015,kandala_hardware-efficient_2017,ryabinkin_qubit_2018,lee_generalized_2019,gard_efficient_2020,meitei_gate-free_2021}.
A popular solution to the ansatz ambiguity problem is the adaptive construction of $\hat{U}$ to iteratively grow the ansatz~\cite{grimsley_adaptive_2019,ryabinkin_iterative_2020,tang_qubit-adapt-vqe_2021,yordanov_qubit-excitation-based_2021,zhang_mutual_2021,zhang_variational_2022}.
These adaptive methods are effective for describing ground states, but their use for excited states has also been investigated~\cite{zhang_adaptive_2021,chan_molecular_2021,yordanov_molecular-excited-state_2022,fitzpatrick_self-consistent_2024}.

\subsection{MORE-ADAPT-VQE}
Here, we outline a generalization of one such strategy, ADAPT-VQE \cite{grimsley_adaptive_2019}, to compute excited states in addition to the ground state based on state averaging.
We will point out the simplifications of the algorithm in the special case of the single-reference version.
We denote the approach Multistate-Objective, Ritz Eigenspectral (MORE-) ADAPT-VQE, because of the two main generalizations in the algorithm:
\begin{enumerate}
	\item An objective function is used based on the weighted energies of an arbitrary number of states. This feature is common to the subspace-search (SS-) \cite{nakanishi_subspace-search_2019} and multistate contracted (MC-) \cite{parrish_quantum_2019} VQEs. 
	\item $\hat{H}$ is diagonalized in the basis of these approximate states to obtain more accurate energies and wavefunctions. This diagonalization scheme is only used in MC-VQE.
\end{enumerate}
When we refer to MORE-ADAPT-VQE, we do not include any orbital optimization effects as in the SA ADAPT-VQE-SCF of Ref.~\citenum{fitzpatrick_self-consistent_2024}.

We begin by choosing a set of $k\in \mathbf{N}$ orthonormal reference states $\{\ket{\phi_i}\}$ with weights $\{\omega_i\}$ such that $\sloppy{\sum_{i=1}^{k} \omega_i = 1}$. 
For the special case of MORE-ADAPT-VQE where $\sloppy{k = 1}$, the algorithm becomes equivalent to the ordinary, state-specific ADAPT-VQE.
We also choose a pool of anti-Hermitian generators $\{\hat{A}_i\}$ to be considered for use in our unitary $\hat{U}\left(\boldsymbol\theta\right)$.
The Hamiltonian $\hat{H}$ is defined by the chemical system under consideration.
For our purposes, the references, generators, and Hamiltonian will be defined in terms of fermionic, second-quantized operators $\{\hat{a}_p\}$ and $\{\hat{a}_p^\dagger\}$.
On real hardware, these would need to be translated into the qubit basis using a mapping such as the Jordan--Wigner transformation~\cite{jordan_uber_1928}.
In this work, the mapping is not relevant and will only be implied throughout.
Finally, we need one or more convergence criteria. These could be based on some metric of state accuracy or on hardware limitations, such as the gradient resolution needed to continue the algorithm or the circuit depth associated with $\hat{U}\left(\boldsymbol\theta\right)$.

MORE-ADAPT-VQE proceeds as follows:

\begin{enumerate}
	\item Define the functions:
		\begin{align}
			E_i \left(\hat{U} \left( \boldsymbol\theta\right)\right) &= \braket{\phi_i|\hat{U}^\dagger\left(\boldsymbol\theta\right)\hat{H}\hat{U}\left(\boldsymbol\theta\right)|\phi_i}\\
			\epsilon_{ij} \left(\hat{U} \left( \boldsymbol\theta\right)\right) &= \braket{\phi_i|\hat{U}^\dagger\left(\boldsymbol\theta\right)e^{-\theta_j\hat{A}_j} \hat{H}e^{\theta_j\hat{A}_j}\hat{U}\left(\boldsymbol\theta\right)|\phi_i}
		\end{align}
	\item Initialize $\hat{U}\left(\boldsymbol\theta\right) = \hat{\mathbf{1}}$, a trivial, unparameterized ansatz, and $\boldsymbol\theta = \left<\right>$, an empty vector which will increase in length throughout the algorithm.
	\item For each generator $\hat{A}_i$, compute a gradient vector $\mathbf{g}$, where
		\begin{align}
			g_j = \frac{\partial}{\partial \theta_j}\sum_i \omega_i \epsilon_{ij} \bigg|_{\theta_j = 0}. 
		\end{align}
	\item Choose the operator $\hat{A}_j$ with the largest gradient $|g_j|$ and append it to the ansatz, updating $\hat{U}$ and $\boldsymbol{\theta}$:
		\begin{align}
			\boldsymbol\theta &\leftarrow \left<0, \boldsymbol\theta\right>\\ 
			\hat{U}\left(\boldsymbol{\theta}\right)&\leftarrow e^{\theta_0 \hat{A}_{l_0}}\hat{U}\left(\boldsymbol\theta\right) 
		\end{align}
		In this case, $l_0 = j$.
		The notation exists to distinguish between the index of an operator within the pool and its position in the ansatz.
		For example, the first operator in the ansatz is not generally $\hat{A}_0$, but some other $\hat{A}_l$.
		As this notation suggests, the algorithm is permitted to add the same operator multiple times as the ansatz grows.

	\item Perform an MC- (\textit{i.e.} state-averaged) VQE subroutine where $\boldsymbol\theta$ is optimized to minimize $E_{\mathrm{SA}}$:
		\begin{align}
			E_{\text{SA}}\left(\boldsymbol\theta\right) &= \sum_i \omega_i E_i\left(\hat{U}\left(\boldsymbol\theta\right)\right)
		\end{align}

		Then, as in MC-VQE, compute the elements of the dressed Hamiltonian $\bar{H}$, where
		\begin{align}
			\bar{H}_{ij} &= \braket{\phi_i|\hat{U}^\dagger\left(\boldsymbol\theta\right)\hat{H}\hat{U}\left(\boldsymbol\theta\right)|\phi_j}.
		\end{align}
		Then diagonalize $\bar{H}$ in the basis of the reference states $\left\{\ket{\phi_i}\right\}$ to obtain the  eigenstates $\left\{\ket{\psi_i}\right\}$:
		\begin{align}
			\bar{H}\ket{\psi_i} &= E_i'\ket{\psi_i} 
		\end{align}
		The resulting eigenstates will be linear combinations of the references with coefficients $\left\{c_{ij}\right\}$:
		\begin{align}
			\ket{\psi_i} = \sum_j c_{ji} \ket{\phi_j}
		\end{align}
		We will define
		\begin{align}
			\ket{\Psi_i} = \hat{U}\ket{\psi_i},
		\end{align}
		such that $\ket{\Psi_i}$ is the actual approximate eigenstate of the original Hamiltonian $\hat{H}$.
	\item If a convergence criterion has not yet been satisfied, return to step 3.  
		Otherwise, return the set of energies $\left\{E_i'\right\}$.  
		Each $E_i'$ is bounded below by the true $i$th lowest energy, a useful feature of the Ritz diagonalization~\cite{hendekovic_energy_1982}.
\end{enumerate}

In traditional ADAPT-VQE, $k = 1$.
This makes step 6 trivial, since 
\begin{align}
	\ket{\Psi_0} = \hat{U}\left(\mathbf{\theta}\right)\ket{\phi_0}. 
\end{align}

Other operators $\hat{O}$ can also be measured in the basis of $\{\hat{U}\ket{\phi_i}\}$, then rotated based on the Ritz eigenvectors of $\hat{H}$ to yield elements $\{\braket{\Psi_i|\hat{O}|\Psi_j}\}$. 
This is useful in computing other properties of the wavefunction, such as the expectation value of $\hat{S}^2$ or the electric dipole operator. 
We point out that MORE-ADAPT-VQE does not necessarily monotonically improve every energy with every iteration. 
While the $i^{\mathrm{th}}$ estimate of the energy is bounded below by the true $i^{\mathrm{th}}$ FCI energy~\cite{hendekovic_energy_1982}, MORE-ADAPT-VQE will increase one energy if it means lowering the average energy, as one would expect of a state-averaged method. 
A real implementation of MORE-ADAPT-VQE would obtain energy and gradient values in steps 3 and 5 by preparing and measuring states on hardware, along with computing the elements of $\bar{H}$. 
In this work, however, these measurements are simulated in the absence of noise or finite shot count error.

\subsection{q-sc-EOM}
We compare our results to the Quantum, Self-Consistent Equation-of-Motion (q-sc-EOM), a powerful quantum algorithm based on expansion about a ground state.~\cite{asthana_quantum_2023}
		In this method, a reference $\ket{\phi_0}$ and a unitary $\hat{U}$ are designated, along with a set of orthonormal references $\{\ket{\phi_{i\neq 0}}\}$. 
		The ground-state wavefunction is assumed to be $\hat{U}\ket{\phi_0}$, and the excited state energies are obtained through a diagonalization of a dressed Hamiltonian 
		\begin{equation}\bar{H} = \hat{U}^\dagger \hat{H} \hat{U}\end{equation}
			in the basis of the other references $\{\ket{\phi_{i\neq 0}}\}$.

		The method is formally and practically attractive: It satisfies the vacuum annihilation (``killer'') condition~\cite{dalgaard_expansion_1979}, is size-intensive, and, as the basis states are orthonormal, does not require noisy overlap measurements to construct the metric  $\braket{\Psi_i|\Psi_j}$ for a generalized eigenvalue problem.
These features combine to yield a strong representative of expansion-based methods.
Additionally, the method was originally designed and has primarily been used with an ansatz obtained through single-reference ADAPT-VQE, making it a natural point of comparison for MORE-ADAPT-VQE.
The q-sc-EOM approach is described in more detail, especially regarding its connection to classical equation-of-motion theory, in Ref.~\citenum{asthana_quantum_2023}.

We use q-sc-EOM in conjunction with traditional ADAPT-VQE to obtain energies for comparison with MORE-ADAPT-VQE.
The expansion manifold of q-sc-EOM is chosen to match that of MORE-ADAPT-VQE in each simulation to achieve a fair comparison , as opposed to a large manifold based on a maximum excitation rank, such as the more commonly used set of singly and doubly excited determinants.
Our  chosen manifolds are  not generally optimal, but we aim to demonstrate that MORE-ADAPT-VQE is able to compensate for the deficiences of its references in a way that an expansion-style correction to ADAPT-VQE cannot.

\section{Numerical simulations}\label{results}
All simulations were performed without noise using the STO-6G basis~\cite{hehre1969a, hehre1970a}.
All ADAPT-VQE calculations were performed with the full set of generalized single and double excitations\cite{nooijen_can_2000}, anti-symmetrized as in unitary coupled cluster\cite{kutzelnigg_w_52_1977}, as the operator pool:
\begin{align}
	\{\hat{A}_i\} = \{\hat{a}_p^q-\hat{a}_q^p\} \cup \{\hat{a}_{pq}^{rs} - \hat{a}_{rs}^{pq}\},
\end{align}
where the indices $p,q,r,s$ run over all fermionic modes.
In the absence of noise, operators that violate $\hat{S}_\mathrm{z}$, particle number, or point group symmetry will never be chosen by ADAPT-VQE for these problems, and are excluded.
The operators do not generally preserve $\hat{S}^2$ symmetry, however.
In all MORE-ADAPT-VQE calculations, equal weights $\frac{1}{k}$ were used for each reference.
We discuss the implications of this choice in section \ref{conclusions}. 
All VQE subroutines were performed using the Broyden--Fletcher--Goldfarb--Shanno (BFGS) algorithm to minimize the objective function~\cite{fletcher_practical_2000}.
FCI and CASSCF results were obtained using \textsc{Psi4}~\cite{smith_psi4_2020}.
All single-reference ADAPT-VQE calculations were allowed to achieve FCI accuracy.  
In a real application, the exactness of the energy is not a valid stopping criterion, since the FCI energy cannot be known \textit{a priori}.
The use of an exact ground state, however, allows us to isolate the behavior of q-sc-EOM from the problem of an inaccurate ADAPT-VQE calculation. 
In MORE-ADAPT-VQE calculations, we report the results obtained by terminating the algorithm at specific numbers of operators being included in the ansatz, specified throughout this section. 
This is similar in spirit to using a circuit depth stopping criterion.

Single-reference ADAPT-VQE, MORE-ADAPT-VQE, and \sloppy{$\text{q-sc-EOM}$} results were obtained with \textsc{QForte}~\cite{stair_qforte_2022}.

\subsection{Rectangular H$_4$ dissociation}
\begin{figure*}
	\includegraphics[width = \textwidth]{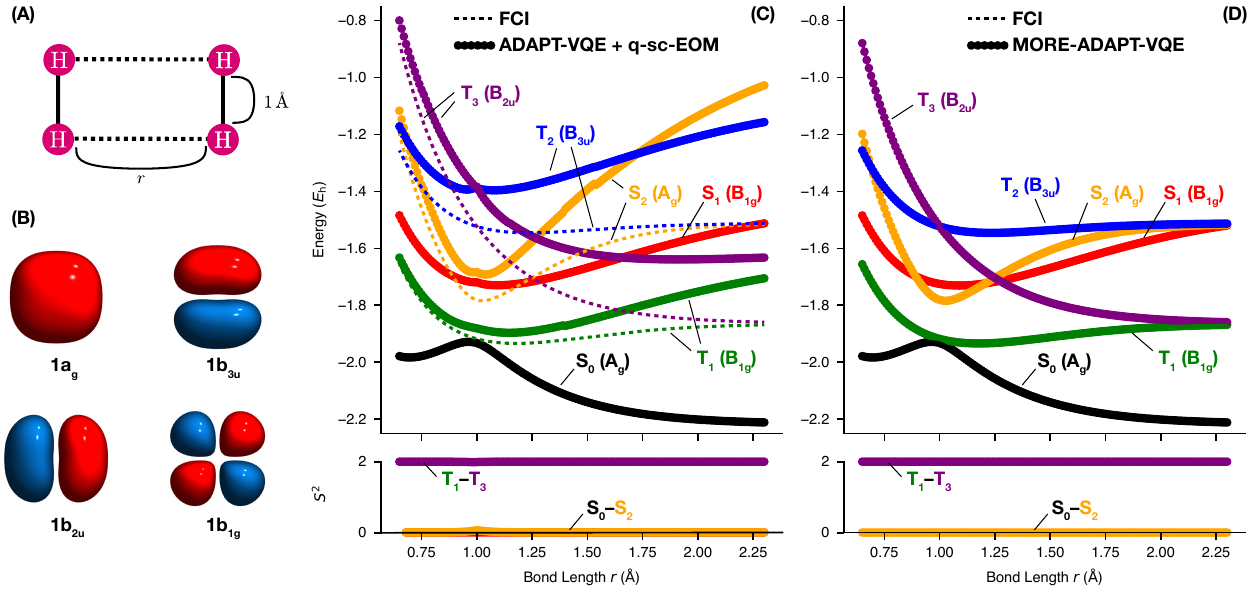}
	\caption{
Rectangular H$_4$ dissociation.
(A) Diagram of the dissociation trajectory. As $r$ grows larger, H$_4$ dissociates into two H$_2$ molecules.
(B) Canonical orbitals of the square H$_4$ geometry ($r = 1$ \AA{}), visualized with the \textsc{VMD} software package~\cite{HUMP96}.
(C) Energy curves of the states of interest obtained with traditional ADAPT-VQE after 11 iterations and followed by q-sc-EOM.
(D) Energy curves of the states of interest obtained with MORE-ADAPT-VQE after 50 iterations.
The different colors indicate the different states.  
Circular markers correspond to the energies obtained with each quantum method, while the dashed lines correspond to the FCI energies. 
In panels (C) and (D), the expectation value of the $\hat{S}^2$ operator of each state is reported at the bottom as a function of $r$. 
}\label{h4_square}
\end{figure*}

We first consider the dissociation of rectangular H$_4$ into two H$_2$ molecules. 
We targeted six low-lying states in analogy with reference~\citenum{smart_many-body_2024}.
The dissociation trajectory from that study was also used, where the H$_2$ molecules have a fixed bond length of 1 $\mathrm{\AA{}}$.
D$_{2\mathrm{h}}$ point group symmetry was used.
The molecular geometry as a function of the H$_2$-H$_2$ distance $r$ is depicted in Panel A of Figure \ref{h4_square}.  
The canonical Hartree--Fock orbitals at the square geometry are shown in Panel B.
This corresponds to A$_{\mathrm{g}}$, B$_{1\mathrm{g}}$, B$_{2\mathrm{u}}$, and B$_{3\mathrm{u}}$ spaces of 12, 8, 8, and 8 determinanants, respectively, for a total relevant Hilbert space of dimension 36.
Canonical orbitals were used for all methods in studying this system. 

Over the course of the rectangular H$_4$ dissociation, the Hartree--Fock determinant switches from 
$\ket{1\mathrm{a}_{\mathrm{g}}^{2}1\mathrm{b}_{2\mathrm{u}}^{2}}$ for $r < 1$ $\mathrm{\AA{}}$ to $\ket{1\mathrm{a}_{\mathrm{g}}^{2}1\mathrm{b}_{3\mathrm{u}}^{2}}$ for $r > 1\text{ }\mathrm{\AA{}}$.
The degeneracy of these determinants at the square D$_{4\mathrm{h}}$ geometry makes rectangular H$_4$ a classic, challenging test system for single-reference methods.~\cite{jankowski_applicability_1980}  
Additionally, the six states considered here have several crossings in their potential energy curves, as well as an avoided crossing between two A$_{\mathrm{g}}$ singlet states at the square geometry, making this an attractive model problem for photochemical applications.

\begin{table}
\caption{CSFs used in the quantum methods for the dissociation of rectangular H$_4$.}
\centering
\footnotesize
{\renewcommand{\arraystretch}{2}
	 \begin{tabular}{cccc}
	 \hline
	 
	 \hline	 	 
	 \# & Spin$^a$ & Symmetry & CSF\\
	 \hline
		1 & 1 & A$_{\mathrm{g}}$ & $\ket{1\mathrm{a}_{\mathrm{g}}^2\mathrm{1b}_{\mathrm{2u}}^2}$ \\
		2 & 1 & A$_{\mathrm{g}}$ & $\ket{1\mathrm{a}_{\mathrm{g}}^2\mathrm{1b}_{\mathrm{3u}}^2}$ \\
		3 & 1 & B$_{1\mathrm{g}}$ & $\frac{1}{\sqrt{2}}\left(\ket{\mathrm{1a}_{\mathrm{g}}^2 \mathrm{1b}_{2\mathrm{u}}^\alpha \mathrm{1b}_{3\mathrm{u}}^\beta} - \ket{\mathrm{1a}_{\mathrm{g}}^2 \mathrm{1b}_{2\mathrm{u}}^\beta \mathrm{1b}_{3\mathrm{u}}^\alpha} \right)$\\
		4 & 3 & B$_{1\mathrm{g}}$ & $\frac{1}{\sqrt{2}}\left(\ket{\mathrm{1a}_{\mathrm{g}}^2 \mathrm{1b}_{2\mathrm{u}}^\alpha \mathrm{1b}_{3\mathrm{u}}^\beta} + \ket{\mathrm{1a}_{\mathrm{g}}^2 \mathrm{1b}_{2\mathrm{u}}^\beta \mathrm{1b}_{3\mathrm{u}}^\alpha} \right)$\\
		5 & 3 & B$_{2\mathrm{u}}$ & $\frac{1}{\sqrt{2}}\left(\ket{\mathrm{1a}_{\mathrm{g}}^\alpha \mathrm{1b}_{2\mathrm{u}}^\beta \mathrm{1b}_{3\mathrm{u}}^2} + \ket{\mathrm{1a}_{\mathrm{g}}^\beta \mathrm{1b}_{2\mathrm{u}}^\alpha \mathrm{1b}_{3\mathrm{u}}^2} \right)$\\
		6 & 3 & B$_{3\mathrm{u}}$ & $\frac{1}{\sqrt{2}}\left(\ket{\mathrm{1a}_{\mathrm{g}}^\alpha \mathrm{1b}_{2\mathrm{u}}^2 \mathrm{1b}_{3\mathrm{u}}^\beta} + \ket{\mathrm{1a}_{\mathrm{g}}^\beta \mathrm{1b}_{2\mathrm{u}}^2 \mathrm{1b}_{3\mathrm{u}}^\alpha} \right)$\\
	 \hline
	 
	 \hline
\end{tabular}

$^a$ Reported as the multiplicity, $2S+1$.
}
	\label{h4_refs}
\end{table}

For both quantum excited state approaches, six configuration state functions (CSFs) were used as the references $\left\{\ket{\phi_i}\right\}$ (see Table \ref{h4_refs}). 
These were chosen due to their dominant contributions to the eigenstates of interest at the square geometry.
For each geometry, we first allowed traditional, single-reference ADAPT-VQE to run on one reference until the FCI energy was recovered (error less than $1\times 10^{-13}$ $E_{\mathrm{h}}$), obtaining an ansatz $\hat{U}$ in the process.
This required 11 ADAPT-VQE iterations.
We then approximated the excited state energies with a q-sc-EOM diagonalization in the subspace of the remaining 5 CSFs. 
We note that in this case, due to the exactness of the ground state, there is no coupling through $\bar{H}$ between the reference and the other CSFs, so that this is equivalent to simpy diagonalizing in the basis of all six CSFs.
Given the degeneracy at \sloppy{$1\text{ }\mathrm{\AA{}}$}, we report q-sc-EOM results for ADAPT-VQE calculations using both the $\ket{1\mathrm{a}_{\mathrm{g}}^{2}1\mathrm{b}_{2\mathrm{u}}^{2}}$ and $\ket{1\mathrm{a}_{\mathrm{g}}^{2}1\mathrm{b}_{3\mathrm{u}}^{2}}$ determinants as references, corresponding to the crossing of the HOMO and LUMO at $r = 1$ $\mathrm{\AA{}}$. In Panel C of Figure \ref{h4_square}, curves from ADAPT-VQE combined with q-sc-EOM are reported, where the determinant of lowest energy is always used as the reference.

We observe several deficiencies of using a unitary $\hat{U}\left(\boldsymbol\theta\right)$ based on a single ADAPT-VQE calculation.  
While ADAPT-VQE was able to recover the ground state FCI energy and wavefunction across the curve, the unitary it used to achieve this does not smoothly vary from one point to the next. 
For example, in Panel C of Figure \ref{h4_square}, using ADAPT-VQE with the $\ket{1\mathrm{a}_{\mathrm{g}}^2 1\mathrm{b}_{\mathrm{3u}}^2}$ reference gave an abrupt increase in energy for the  S$_{\mathrm{2}}$ energy curve at $1.44$ $\mathrm{\AA{}}$.
This corresponds to a change in the fourth operator of the ansatz, which switches from being the HOMO-LUMO pair excitation to the $\hat{a}_{0\bar{1}}^{3\bar{2}}$ excitation.
While both operator sequences can yield the correct ground-state wavefunction, the two distinct unitaries act differently on the other references, giving discontinuities in the excited state energies, regardless of the accuracy of the ground state. 
Aside from generating jagged potential energy surfaces, q-sc-EOM failed to capture basic qualitative features of the H$_4$ dissociation.
Most conspicuously, the excited A$_{\mathrm{g}}$ singlet fails to cross the B$_{\mathrm{1g}}$ singlet, and an incorrect crossing between the excited A$_{\mathrm{g}}$ singlet and the B$_{\mathrm{3u}}$ triplet is present.
Additionally, q-sc-EOM failed to even approximately describe the degeneracies of the different states in the limit of complete dissociation. Without the introduction of a larger set of references, the combination of ADAPT-VQE and q-sc-EOM cannot improve any further, since ADAPT-VQE itself has already converged.

We report MORE-ADAPT-VQE results for the rectangular H$_4$ dissociation based on the same set of references shown in Table \ref{h4_refs}.
We did not privilege any of the references. 
The resulting energy curves are pictured in Panel D of Figure \ref{h4_square}.
50 operators were used, by which point all of the FCI energies had been recovered (error less than $1\times 10^{-10}$ $E_{\mathrm{h}}$ for each state).
In obtaining these exact energies, we demonstrate that a fundamental limitation of a single-reference ADAPT-VQE calculation can be addressed by tailoring the ansatz itself to target multiple eigenstates of $\hat{H}$.

This result is fairly surprising.  
While there necessarily exists a $\hat{U}$ that block-diagonalizes $\hat{H}$ to isolate a block containing the targeted excited state spectrum, it is not obvious that this $\hat{U}$ could be prepared with the UCCGSD ansatz, let alone discovered by a greedy search algorithm like MORE-ADAPT-VQE.
We discuss this further in section \ref{conclusions}.
While H$_4$ is a small system with solutions of mostly different symmetries, we  also see excellent agreement between MORE-ADAPT-VQE and FCI for more states with shared symmetries in the case of linear H$_4$, as well as for larger, 7-orbital BeH$_2$ calculations.


\subsection{Linear H$_4$ dissociation}

\begin{figure*}
	\centering
		\includegraphics[width = \textwidth]{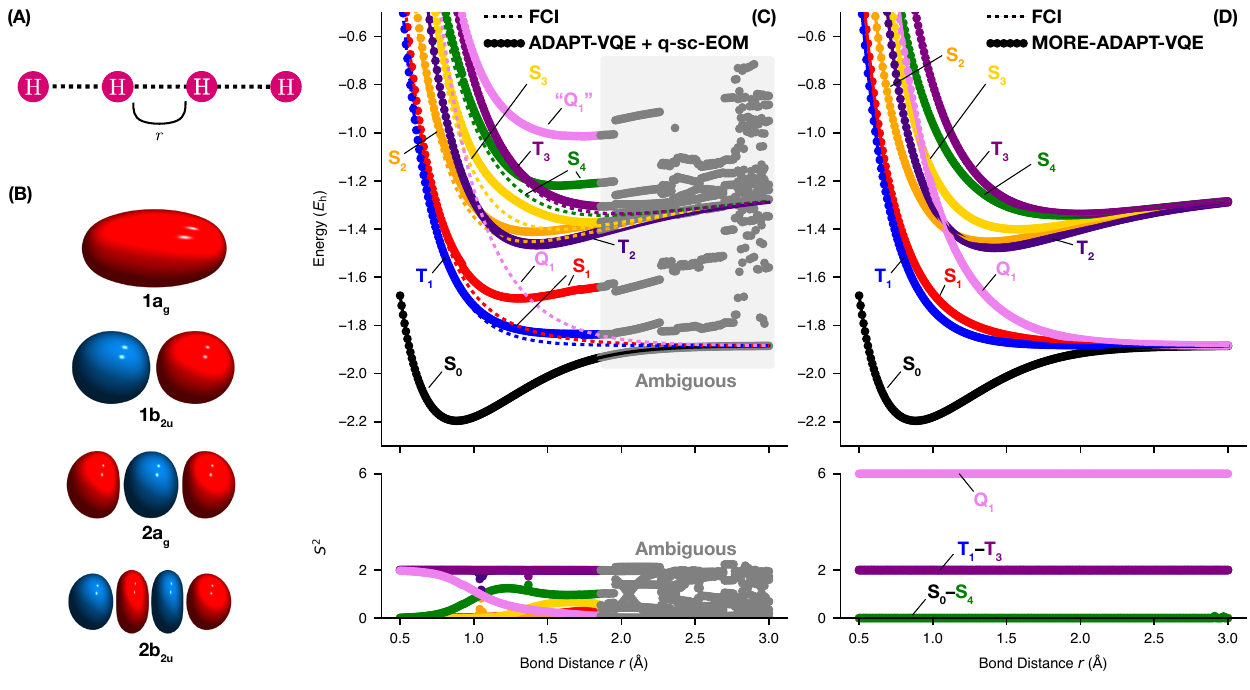}
	\caption{
Linear H$_4$ dissociation.
(A) Diagram of the dissociation trajectory. As $r$ grows larger, H$_4$ dissociates into four hydrogen atoms.
(B) Canonical orbitals of the equilibrium ($r = 0.88$ \AA{}) linear H$_4$ geometry, visualized with the \textsc{VMD} software package~\cite{HUMP96}.
(C) Energy curves of the states of interest obtained with traditional ADAPT-VQE after 20 iterations and followed by q-sc-EOM.
	(D) Energy curves of the states of interest obtained with MORE-ADAPT-VQE after 100 iterations.
The different colors indicate the different states.  Circular markers correspond to the energies obtained with each quantum method, while the dashed lines correspond to the FCI energies. 
In panels (C) and (D), the expectation value of the $\hat{S}^2$ operator of each state is reported at the bottom as a function of $r$.
In Panel (C), the assignment of the states from q-sc-EOM is challenging and becomes ambiguous after $r = 1.8$ \AA{}, as is indicated by the grey region of the plot.
The pink curve labeled ``Q$_1$'' (\textit{sic}) does not correspond to a true quintet, but rather the state which is found instead of the target quintet.}
\label{linear_h4}
\end{figure*}

Linear hydrogen chains are commonly used as prototypes for strongly correlated molecules, including for ADAPT-VQE~\cite{jankowski_applicability_1980, grimsley_adaptive_2019}.
We consider the specific case of an equally spaced chain of four hydrogens, as shown in Panel A of Figure \ref{linear_h4}. 
The system's canonical Hartree--Fock orbitals at the equilibrium geometry ($r = 0.88$ $\mathrm{\AA{}}$) are shown in Panel B of Figure \ref{linear_h4}.
The canonical orbitals were used in all methods. 
For this system, we targeted only A$_\mathrm{g}$ states using  D$_{2\mathrm{h}}$ point group symmetry, giving a Hilbert space of  20 determinants.

We consider here the nine states of A$_{\mathrm{g}}$ symmetry which are lowest in energy.
There are several crossings between excited states near the equilibrium geometry, notably between S$_4$ and Q$_1$ at 0.75 $\mathrm{\AA{}}$, S$_3$ and Q$_1$ at 0.92 $\mathrm{\AA{}}$, T$_2$ and Q$_1$ at 1.07 $\mathrm{\AA{}}$, S$_2$ and Q$_1$ at 1.09 $\mathrm{\AA{}}$, and S$_2$ and T$_2$ at 1.15 $\mathrm{\AA{}}$, providing several degeneracies between states of the same point group symmetry.  
While these crossings are allowed due to $\hat{S}^2$ symmetry, effective Hamiltonian constructions $\bar{H}$ from q-sc-EOM or MORE-ADAPT-VQE will not generally have spin-pure eigenstates, possibly preventing these crossings.
While this is also possible for the other two systems considered in this work, spin-contamination was more significant here, exacerbated by using individual determinants as references rather than CSFs, providing an interesting test case.

For both quantum methods, nine individual Slater determinants of A$_{\mathrm{g}}$ symmetry were used as references $\left\{\ket{\phi_i}\right\}$ based on their large contributions to the exact eigenstates at the equilibrium geometry (0.88 $\mathrm{\AA{}}$): $\ket{1\mathrm{a}_{\mathrm{g}}^2 1\mathrm{b}_{2\mathrm{u}}^2}$, $\ket{1\mathrm{a}_{\mathrm{g}}^2 2\mathrm{a}_{\mathrm{g}}^2}$, $\ket{1\mathrm{b}_{2\mathrm{u}}^2 2\mathrm{a}_{\mathrm{g}}^2}$, $\ket{1\mathrm{a}_{\mathrm{g}}^2 1\mathrm{b}_{2\mathrm{u}}^\alpha 2\mathrm{b}_{2\mathrm{u}}^\beta}$,
$\ket{1\mathrm{a}_{\mathrm{g}}^{\alpha} 1\mathrm{b}_{2\mathrm{u}}^2 2\mathrm{a}_{g}^\beta}$, and
$\ket{1\mathrm{a}_{\mathrm{g}}^{\alpha} 1\mathrm{b}_{2\mathrm{u}}^\alpha 2\mathrm{a}_{\mathrm{g}}^\beta 2\mathrm{b}_{2\mathrm{u}}^\beta}$, as well as the spin complements of the latter three determinants.

For each geometry, a single-reference ADAPT-VQE calculation was run to completion based on the $\ket{1\mathrm{a}_{\mathrm{g}}^2 1\mathrm{b}_{2\mathrm{u}}^2}$ determinant (error less than $1\times 10^{-8}$ $E_{\mathrm{h}}$). 
20 operators were needed to exactly reproduce the FCI ground state energy curve.  The remaining 8 determinants were then used as an expansion manifold for q-sc-EOM.  The resulting ground and excited states are shown in Panel C of Figure \ref{linear_h4}. 

In the case of linear H$_4$, q-sc-EOM struggled even more with recovering the excited states based on single-reference ADAPT-VQE calculations.
In Panel C of Figure \ref{linear_h4}, we see several major, qualitative issues.
Severe spin-contamination is observed among the resulting curves, even near equilibrium, along with a complete failure to capture the quintet across the curve.
Furthermore, a spurious crossing between S$_4$ and T$_3$ (which should be the two highest excited states to qualitatively match FCI) is observed at $r = 1.37$ $\mathrm{\AA{}}$.
As the molecule dissociates, the q-sc-EOM states become increasingly jagged, becoming impossible to assign to the FCI states they should be approximating.

MORE-ADAPT-VQE calculations were run using all nine determinants as equally weighted references.
The resulting energy curves from 100 iterations of MORE-ADAPT-VQE are shown in Panel D of Figure \ref{linear_h4}, by which point most of the MORE-ADAPT-VQE calculations had obtained the exact FCI energies for all nine states (error less than $1\times 10^{-10}$ $E_{\mathrm{h}}$ for each state).

We again see that traditional ADAPT-VQE was unable to obtain a good unitary operator such that $\bar{H}$ had the desired spectrum, no matter how many iterations it ran, failing to accurately capture even S$_1$ and T$_1$.
This is particularly troubling given that almost half the Hilbert space (accounting for particle number, $M_s$, and point group symmetries) was included in the q-sc-EOM diagonalization.
MORE-ADAPT-VQE, however, was able to recover excellent energy curves by assembling its ansatz based on all of the references.

Of particular interest is the quintet state.  
While the reference determinants do not span the full space of $M_s = 0$ determinants that form the quintet CSFs, MORE-ADAPT-VQE was able to capture this state.
Even though the operator pool is not spin symmetry-preserving, traditional ADAPT-VQE has no incentive to break the spin symmetry of the excited states spanned by the references.
For states with significant open-shell character with $M_s = 0$,  important  CSFs will consist of linear combinations of many determinants.
The greater  the number of determinants in the references, the more expensive it becomes to compute off-diagonal elements of $\bar{H}$, resulting in potentially large savings in quantum resources if MORE-ADAPT-VQE is used with single-determinant references.
Given the severity of spin-contamination in the q-sc-EOM results for linear H$_4$ relative to the q-sc-EOM results for the other two systems, which used CSFs instead, using individual determinants is a poor idea for small-manifold q-sc-EOM.
The comparatively forgiving nature of MORE-ADAPT-VQE in dealing with spin-contaminated references is a very attractive feature.

\subsection{Insertion of beryllium into H$_2$}
\begin{figure*}
	\centering
	\includegraphics[width = \textwidth]{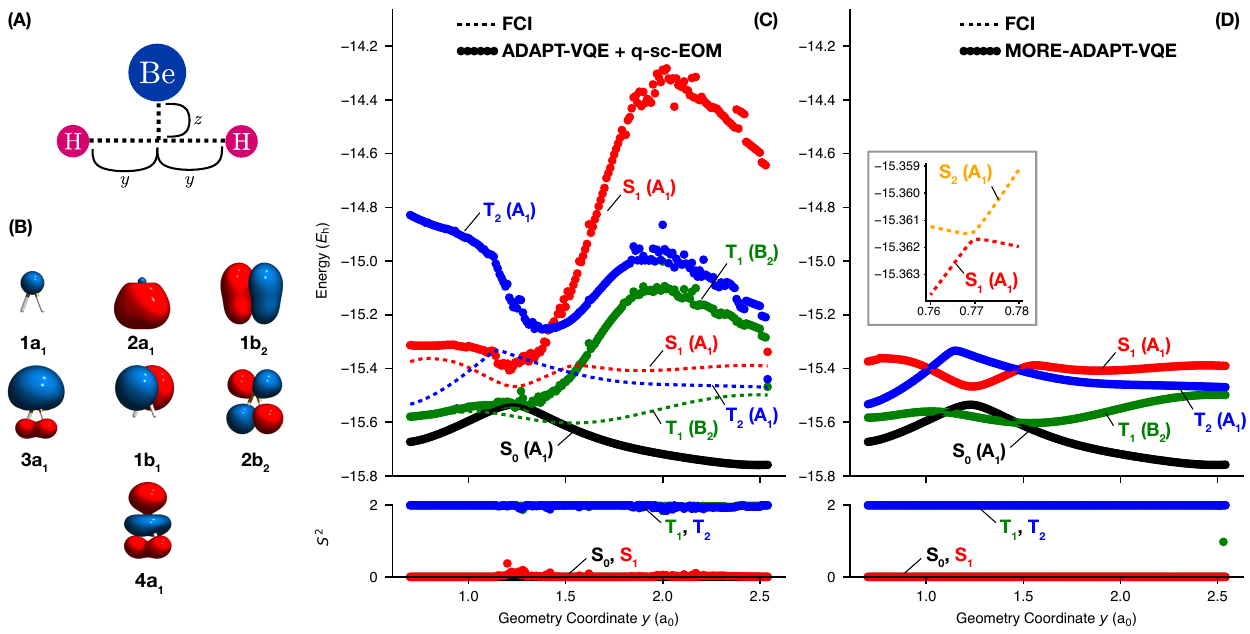}
	\caption{
Be insertion into H$_2$.
(A) Diagram of the BeH$_2$ insertion trajectory, where $z = \frac{127 - 50y}{23}$ $\mathrm{a}_0$. 
(B) CASSCF orbitals at the BeH$_2$ transition state geometry ($y = 1.23$ a$_0$), visualized with the \textsc{VMD} software package~\cite{HUMP96}.
(C) Energy curves of the states of interest in the dissociation of BeH$_2$ obtained with traditional ADAPT-VQE after 225 iterations and followed by q-sc-EOM.
(D) Energy curves of the states of interest obtained with MORE-ADAPT-VQE after 300 iterations.
The different colors indicate the different states.  Circular markers correspond to the energies obtained with each quantum method, while the dashed lines correspond to the FCI energies.
The inset depicts the S$_1$-S$_2$ avoided crossing.
In panels (C) and (D), the expectation value of the $\hat{S}^2$ operator of each state is reported at the bottom as a function of $r$.
The visible outlier in the T$_1$ energy curve in Panel D corresponds to an extremely low overlap between CSF 4 and the T$_1$ solution and is corrected with additional MORE-ADAPT-VQE iterations.
	}\label{beh2}
\end{figure*}

As a larger model, we consider the insertion of beryllium into an H$_2$ molecule, using a trajectory from reference~\citenum{purvis_iii_c_1983} in the C$_{\mathrm{2v}}$ point group.
We show the insertion pathway in Panel A of Figure \ref{beh2}.
This insertion is a common case study for excited state methods~\cite{chaudhuri_comparison_1997,jimenez-hoyos_excited_2013}.
The S$_0$-S$_1$ avoided crossing at the transition state ($y = 1.23$ a$_0$) is particularly challenging to treat as the determinant composition of the S$_0$ state changes along the geometry coordinate.
For this reason, in all calculations, molecular orbitals were obtained from a state-specific CASSCF calculation with two electrons distributed in the 1b$_2$ and 3a$_1$ active orbitals.
The optimized CASSCF orbitals for the S$_0$ transition state geometry ($y = 1.23$ a$_0$) are shown in Panel B of Figure \ref{beh2}.
Computations based on quantum methods were performed using these orbitals in the full (6,7) active space. 

The individual A$_1$ and B$_2$ subspaces contain 321 and 304 determinants, respectively, for a total relevant Hilbert space of 625 determinants.
Note that CASSCF is used only to choose a better set of orbitals prior to the quantum algorithms.
The orbitals are fixed throughout MORE-ADAPT-VQE, unlike in reference \cite{fitzpatrick_self-consistent_2024}.

We consider the two singlets of A$_1$ symmetry as well as the triplets of A$_1$ and B$_2$ symmetry which are lowest in energy.
At the FCI level, there are two realized crossings between S$_0$ and T$_1$,  two realized crossings between S$_1$ and T$_2$, and an avoided crossing between S$_0$ and S$_1$ at the transition state geometry. 

To obtain the four energy curves for the BeH$_2$ insertion, four  CSFs were used as the references $\left\{\ket{\phi_i}\right\}$, shown in Table \ref{beh2_refs}. 
These CSFs were chosen based on their large contributions to the FCI wavefunctions at the transition state geometry.
A minor technical complication must be addressed for the linear ($y = 2.54$ a$_0$) geometry.
In the linear molecule, the 2p$_{\mathrm{x}}$ and 2p$_{\mathrm{z}}$ orbitals of beryllium can no longer mix with any others.
Consequently, CSF 2 is no longer of the same irreducible representation as the desired excited state singlet, and was replaced by a B$_{1\mathrm{u}}$ CSF denoted as $2^*$, where D$_{2\mathrm{h}}$ symmetry was utilized. 
This also applies to the MORE-ADAPT-VQE simulations.
This results in solutions of A$_{\mathrm{g}}$, B$_{1\mathrm{u}}$, and B$_{3\mathrm{g}}$ symmetry, with Hilbert space dimensions of 169, 152, and 148, respectively, for a total relevant Hilbert space dimension of 469.

\begin{table}
\caption{CSFs used in the quantum methods for the BeH$_2$ insertion.}
\centering
\footnotesize
\renewcommand{\arraystretch}{2}
\begin{tabular}{cccc}
\hline

\hline
\# & Spin$^a$ & Symmetry & CSF \\
\hline
1 & 1 & A$_1$ & $\ket{1\mathrm{a}_1^2 2\mathrm{a}_1^2 1\mathrm{b}_2^2}$ \\
2 & 1 & A$_1$ & $\ket{1\mathrm{a}_1^2 2\mathrm{a}_1^2 3\mathrm{a}_1^2}$ \\
3 & 3 & A$_1$ & $\frac{1}{\sqrt{2}}\left(\ket{1\mathrm{a}_1^2 2\mathrm{a}_1^\alpha 3\mathrm{a}_1^\beta 1\mathrm{b}_2^2} + \ket{1\mathrm{a}_1^2 2\mathrm{a}_1^\beta 3\mathrm{a}_1^\alpha 1 \mathrm{b}_2^2}\right)$ \\
4 & 3 & B$_2$ & $\frac{1}{\sqrt{2}}\left(\ket{1\mathrm{a}_1^2 2\mathrm{a}_1^2 3\mathrm{a}_1^\alpha 1\mathrm{b}_2^\beta} + \ket{1\mathrm{a}_1^2 2\mathrm{a}_1^2 3\mathrm{a}_1^\beta 1\mathrm{b}_2^\alpha}\right)$ \\
2$^*$ & 1 & B$_{\mathrm{1\mathrm{u}}}$ & $\frac{1}{\sqrt{2}}\left(\ket{1\mathrm{a}_{\mathrm{g}}^2 2\mathrm{a}_{\mathrm{g}}^\alpha 1\mathrm{b}_{2\mathrm{u}}^2 1\mathrm{b}_{1\mathrm{u}}^\beta} - \ket{1\mathrm{a}_{\mathrm{g}}^2 2\mathrm{a}_{\mathrm{g}}^\beta 1\mathrm{b}_{2\mathrm{u}}^\beta 1\mathrm{b}_{1\mathrm{u}}^\alpha}\right)$ \\
\hline

\hline
\end{tabular}
\label{beh2_refs}

$^a$ Reported as the multiplicity, $2S+1$.
\end{table}

A singlet ground state energy curve was obtained by running traditional ADAPT-VQE until convergence (error less than $1\times 10^{-7}$ $E_{\mathrm{h}}$) based on CSF 1 from Table \ref{beh2_refs}. 
ADAPT-VQE was able to obtain the exact ground state for all points with 225 operators.  
The other three CSFs were then used as an expansion basis for q-sc-EOM.
The resulting curves are shown in Panel C of Figure \ref{beh2}. 

For the BeH$_2$ insertion, q-sc-EOM does a poor job of describing all three excited states throughout the curve. 
As shown in Panel C of Figure \ref{beh2}, only one of the four realized crossings was even qualitatively predicted, and massive quantitative errors are present throughout the insertion, on the order of hundreds of m$E_{\mathrm{h}}$.

MORE-ADAPT-VQE was performed using all four CSFs as references.
The energy curves obtained using 300 operators are shown in Panel D of Figure \ref{beh2}.
Unlike q-sc-EOM, MORE-ADAPT-VQE was able to efficiently recover very reasonable curves using this small set of references. 
By 300 iterations of MORE-ADAPT-VQE, all four energy curves were qualitatively correct, though there is a slight outlier at 2.53 $\mathrm{\AA{}}$. 
This problematic point on the T$_1$ energy curve is eventually corrected through additional iterations of MORE-ADAPT-VQE, and corresponds to a very small overlap between CSF 4 and the target triplet.
There was also a slight discontinuity in the S$_1$ energy curve for small values of $y$ ($\approx 0.77$~a$_0$), corresponding to an avoided crossing between S$_1$ and a third A$_1$ state, S$_2$, at that point.
The avoided crossing is depicted as an inset in Figure~\ref{beh2}.
This outlier can also be corrected if additional ADAPT-VQE iterations are allowed.

\begin{figure*}
	\centering
	\includegraphics[width = 0.525\textwidth]{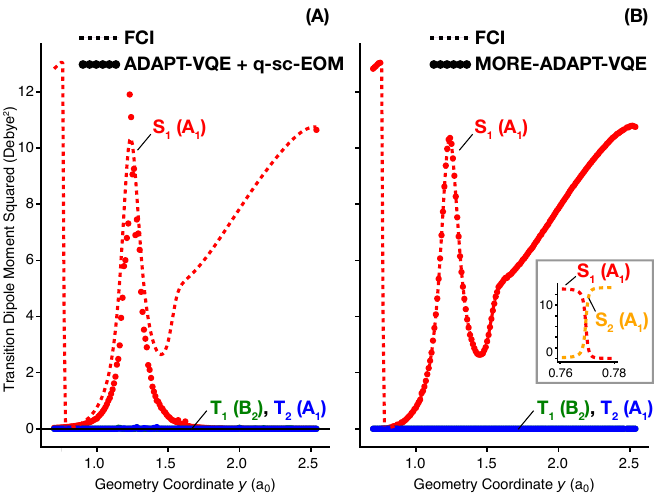}
	\caption{
Be insertion into H$_2$.
		(A): Transition dipole moments squared obtained using the q-sc-EOM solutions from Figure \ref{beh2}.
		(B): Transition dipole moments squared obtained using the MORE-ADAPT-VQE solutions from Figure \ref{beh2}.
		Circular markers correspond to the values obtained with each quantum method, while the dashed lines correspond to the FCI values. 
		The geometry coordinates are the same as those in Figure \ref{beh2}.
		The S$_0$-T$_1$ and S$_0$-T$_2$ transitions have zero transition dipole moment for exact calculations.
		For both quantum methods, the squares of these transition dipole moments were small enough to be largely indiscernible from the x-axis in the figure.
		The apparent discontinuity in the S$_0$-S$_1$ curve is due to an avoided crossing between S$_1$ and a higher singlet, S$_2$, as shown in the inset.}
	\label{beh2_dipoles}
\end{figure*}

In addition to the excited-state energies, the squares of the transition dipole moments between the ground state and excited states were computed based on the q-sc-EOM and MORE-ADAPT-VQE solutions.
For a solution $\ket{\Psi_i}$, the square of the transition dipole moment was computed as
\begin{equation}
	|\mu_{0i}|^2 = |\bra{\Psi_0}\hat{\mu}\ket{\Psi_i}|^2,
\end{equation}
where $\hat{\mu}$ is the electric dipole moment operator.
Such transition dipole moments are related to the probability of a system changing from one state to another with time, such as the excitations induced by light which drive photochemical reactions.
The squares of the dipole moments obtained with the q-sc-EOM and MORE-ADAPT-VQE solutions are shown in Panels A and B, respectively, of Figure \ref{beh2_dipoles}.

For symmetry-forbidden singlet-triplet excitations, q-sc-EOM predicts small but nonzero transition dipole moments, reflecting spin contamination in the excited state solutions.
MORE-ADAPT-VQE also predicts small, nonzero values of $|\mu_{0i}|^2$ for the triplet state, but these are one to two orders of magnitude smaller than the q-sc-EOM values.
These spin-forbidden transitions are analyzed statistically in Table \ref{triplet_int}.
\begin{table}
\caption{BeH$_2$ Singlet-Triplet Transition Dipole Moments Squared (in Debye$^2$)}
\centering
\footnotesize
{\renewcommand{\arraystretch}{2}
	 \begin{tabular}{lcc}
	 \hline
	 
	 \hline
		 {} & q-sc-EOM & MORE-ADAPT-VQE\\
	 \hline
		  S$_0$-T$_1$ (Maximum)& $5.576\times 10^{-2}$ & $6.460\times 10^{-4} $\\
	 \hline
		  S$_0$-T$_1$ (Mean)   & $2.025\times 10^{-3}$ & $2.058\times 10^{-5}$ \\
	 \hline
		  S$_0$-T$_2$ (Maximum) &  $6.069\times 10^{-2}$ & $1.606\times 10^{-3} $\\
	 \hline
		  S$_0$-T$_2$ (Mean) & $1.956\times 10^{-3}$ & $4.514\times 10^{-5}$ \\
	 \hline
	 
	 \hline	 
	 \end{tabular}
	\label{triplet_int}}
\end{table}
Of more interest is the singlet-singlet transition.
In Figure \ref{beh2_dipoles}, we see three key problems with the calculation of this transition dipole moment with q-sc-EOM.
The most obvious issue, consistent with the severe problems in computing the energy of the S$_1$ state in Figure \ref{beh2}, is the incorrect behavior of the method as the molecule becomes more linear.
Here, q-sc-EOM fails to recognize a region of very large transition dipole moments due to a poor description of the S$_1$ state.
As mentioned previously, the references were chosen to represent the excited states of BeH$_2$ at the avoided crossing.
Due to ADAPT-VQE's agnosticism toward the excited state energies, q-sc-EOM fails to obtain good results when the reference manifold becomes less appropriate, even when allowed to run to completion.
A similar problem occurs on the left-hand side of the curve, where there is an avoided crossing between S$_1$ and S$_2$, resulting in a relatively small change in the energies but a large change in the wavefunctions.
This is an illustrative example.
While ADAPT-VQE does not give perfectly smooth curves due to the complicated, discrete optimization involved, it makes intuitive sense that for a ground state that does not rapidly change, the unitary $\hat{U}$ should be a slowly varying function of the geometry.
Given that the same references are used, this should result in an effective Hamiltonian that does not undergo a rapid change during an avoided crossing between excited states.
Because S$_2$ is not well-represented in the effective Hamiltonian throughout the rest of the curve, one would not expect it to abruptly become well-represented just because it becomes lower in energy than S$_1$.
This problem is mitigated by MORE-ADAPT-VQE, which is incentivized to rapidly change the character of its excited state solutions, even when the ground state solution does not change.
We finally note that, even ignoring the jagged curve, the peak transition dipole moment about the transition state geometry is significantly overestimated by q-sc-EOM, an unsurprising result due to the inadequate characterization of S$_1$.

The relative performances of q-sc-EOM and MORE-ADAPT-VQE here are consistent with the two H$_4$ geometries.
This is a more chemically relevant model, however, with the number of references being orders of magnitude smaller than the Hilbert space, and MORE-ADAPT-VQE not being run to completion.
These results offer evidence that MORE-ADAPT-VQE will still be useful for both energies and wavefunction properties of systems of real interest, where large sets of references and exact ansatz preparation are not feasible.

\begin{figure*}
        \centering
        \includegraphics[width = \textwidth]{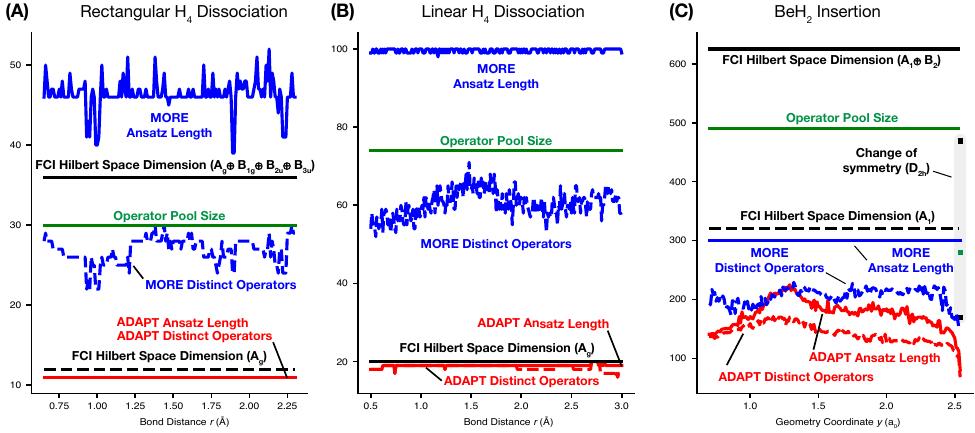}
        \caption{
		Ans{\"a}tze information from numerical simulations of (A) rectangular H$_4$ dissociation, (B) linear H$_4$ dissociation, and (C) BeH$_2$ insertion. 
The geometry coordinates are the same as those in Figures \ref{h4_square}-\ref{beh2}.
The dashed black lines correspond to the totally symmetric FCI Hilbert space dimension associated with the traditional ADAPT-VQE solution, while the solid black lines correspond to the FCI Hilbert space containing all irreducible representations associated with the set of target states for each system. 
	Note that in the case of linear H$_4$, only A$_{\mathrm{g}}$ states were targeted, so that the two FCI spaces are identical.
	The green lines correspond to the number of operators in the UCCGSD pool for each system.
	Recall that the UCCGSD pool does not include operators which are not totally symmetric, since they cannot be chosen by the ADAPT-VQE procedure.
	It follows that rectangular and linear H$_4$ have different UCCGSD pools.
Note that both the FCI spaces and the operator pool decrease in size as BeH$_2$ becomes linear at $y = 2.54$ $\mathrm{\AA{}}$ in panel (C), as indicated by the grey shaded region.
	At that point, the target space includes the A$_{\mathrm{g}}$, B$_{1\mathrm{u}}$, and B$_{3\mathrm{u}}$ irreducible representations.
The red and blue lines correspond to the single-reference ADAPT-VQE calculation needed for q-sc-EOM and the MORE-ADAPT-VQE calculation, respectively.
The solid red and blue lines indicate the total number of operators used in each ansatz, while the dashed lines indicate the number of distinct operators used.
        }
        \label{op_summary}
\end{figure*}

For completeness, we include in Figure \ref{op_summary} some information about the individual ans{\"a}tze constructed for each system at each geometry.
We note a few interesting trends.
First, the number of traditional ADAPT-VQE iterations seems to be approximately bounded by the dimension of the size of the symmetric irreducible representation, though for the larger BeH$_2$ system, far fewer operators are required.
MORE-ADAPT-VQE needs an ansatz $\hat{U}$ which rotates $k$ references into $k$ FCI wavefunctions in Hilbert spaces of sizes $|\mathcal{H}_1|, |\mathcal{H}_2|, \dots,\text{ and }|\mathcal{H}_k|$, respectively.
One might expect that the parameters required for MORE-ADAPT-VQE would be approximately bounded by the sum of the Hilbert space dimensions.
For example, the rectangular H$_4$ calculations in this work would correspond to
\begin{equation}
	\sum_i |\mathcal{H}_i| = 12 + 12 + 8 + 8 + 8 + 8  = 56\text{ determinants}.
\end{equation}
MORE-ADAPT-VQE, using only 50 operators, is within this bound.
Similarly, linear H$_4$ would be expected to require roughly 180 operators for its 9 totally symmetric solutions.
This is far larger than the 100 operators which give all 9 exact FCI states for most bond distances.
This suggests that for large numbers of references with the same irreducible representation, the number of required operators scales significantly better than simply multiplying the size of the irreducible representation by $k$.
If one is not concerned with balanced treatment of solutions of different irreducible representations, the set of references can always be divided up into separate MORE-ADAPT-VQE calculations of distinct irreducible representations.
Further investigation of this idea will be reserved for future work.
Based on the proposed linear upper bound, BeH$_2$ would be expected to require 1267 operators (aside from the linear geometry at $y = 2.54$ a$_0$).
We see that excellent results are obtained with only 300 operators, indicating that the large upper bound will be a gross over-estimation in larger systems.
With respect to the additional operators used by MORE-ADAPT-VQE, it is clear that a larger fraction of the operator pool must be used relative to traditional ADAPT-VQE.
That is, the MORE-ADAPT-VQE ans{\"a}tze are not simply applying the same operators additional times, even in the case of linear H$_4$ where the excited states are of the same spatial symmetry as the ground state.
These results are encouraging and suggest that further studies of larger systems are needed to determine the general viability of the MORE-ADAPT-VQE method in terms of ansatz length.

\section{Conclusions}\label{conclusions}
Describing electronic excitations requires a balanced treatment of both ground and excited states. Expansions based on single-reference wavefunctions often struggle with this, particularly for strongly correlated states. We explore an alternative approach by preparing multiple states simultaneously in a generalized ADAPT-VQE framework.
In our calculations, traditional ADAPT-VQE reliably recovered the exact ground state. However, q-sc-EOM failed to produce qualitatively correct energy curves for any system with a small set of references, while MORE-ADAPT-VQE succeeded using the same reference CSFs or determinants.
Single-reference ADAPT-VQE stops once the ground state is reached, requiring a complete restart in a variational quantum deflation scheme to target excited states~\cite{chan_molecular_2021}.
State-averaged ansatz construction improves the whole set of states simultaneously, obviating the need for many non-parallelizable ADAPT-VQE calculations.

To obtain $k$ different states, MORE-ADAPT-VQE will na{\"i}vely need to perform $k$ times more measurements than traditional ADAPT-VQE at each iteration, since the $k^2$ elements of $\bar{H}$ need only be measured at the very end of the algorithm.
This assumes that the size of $k$ is smaller than the size of the pool, which is a realistic expectation.
In the operator addition step of ADAPT-VQE, a gradient element must be computed for every element in the pool.
In the UCCGSD pool, for example, this corresponds to $\mathcal{O}\left(N^4\right)$ gradients, with an additional factor of $k$ from the different references in MORE-ADAPT-VQE, while $\bar{H}$ only has $k^2$ elements to measure.
For intensive properties like excitation energies, $k$ can be constant with respect to system size.
We have seen in section \ref{results} that MORE-ADAPT-VQE does not seem to require more than $k$ references to obtain these $k$ excited states.
This would imply that the dimension of $\bar{H}$ does not need to increase just because the Hilbert space becomes combinatorially larger with respect to system size.
Our results suggest that this is not the case for q-sc-EOM, however. If the full set of singles and doubles used in the original paper\cite{asthana_quantum_2023} is needed as an expansion basis, the construction of $\bar{H}$ will require $\mathcal{O}\left(N^8\right)$ measurements as a function of system size $N$.
Testing on larger systems is needed to meaningfully compare the necessary sizes of the expansion manifolds for each method, but the current data suggests that MORE-ADAPT-VQE achieves better accuracy with the same number of references. 
Like q-sc-EOM, MORE-ADAPT-VQE has the benefit of not computing an overlap matrix, which has been shown to substantially improve resilience to noise. 
Furthermore, it requires no additional qubits compared to ground state methods, provided linear combinations of reference pairs are easily prepared, utilizing the off-diagonal measurement scheme from Ref.~\citenum{asthana_quantum_2023}.
Although q-sc-EOM fails to recover correct energy curves with small excitation manifolds, even with a converged ground state, it is unclear to what extent MORE-ADAPT-VQE will require longer circuits, which could prove to be a limiting factor.

A formal aspect of state-averaged VQE methods that has not received enough attention is the question of ansatz exactness.
That is, what type of operator pool is sufficient to reproduce exactly the target $k$ eigenstates?
It has been shown that a generalized UCCSD pool of operators can be used to construct a unitary that transforms any reference state into any final state~\cite{evangelista_exact_2019}.
This does not rule out the possibility of ADAPT-VQE finding a local \emph{ansatz minimum} such that the gradient associated with adding each operator in the UCCSD pool is zero, even if the current wavefunction is not an eigenstate of the Hamiltonian.
For MORE-ADAPT-VQE, the question of exact state construction generalizes in the following way: Can a unitary $\hat{U}$ be constructed from a given pool of operators such that
\begin{equation}
	\text{Span}\{\hat{U}\ket{\phi_i}\} = \text{Span}\{\ket{\Psi_i}\}
\end{equation}
for any number of orthogonal references $\{\ket{\phi_i}\}$ and orthogonal target states $\{\ket{\Psi_i}\}$?
If this condition is satisfied, it follows that MORE-ADAPT-VQE can, in the absence of ansatz minima, obtain the exact eigenstates of the Hamiltonian. 
A stronger requirement is that
\begin{equation}
	\hat{U}\ket{\phi_i} = \ket{\Psi_i}
\end{equation}
for all $i$, with the result that properties of individual states would no longer require off-diagonal measurements of operators, such as the electric dipole moment.
 Addressing these conjectures would offer significant insight into the potential and limitations of MORE-ADAPT-VQE.

In this work, we have exclusively used the na{\"i}ve choice of equal weights.
References \cite{ding_ground_2024} and \cite{hong_refining_2024} examine state-averaged cost functions with different weight schemes, as in SS-VQE, providing theoretical justification for using specific sets of weights depending on what cost function one is trying to minimize.
For example, different sets of weights are expected to give better estimates of the wavefunctions than the weights that give the best estimates of the energies.
Using alternative weights merits exploration in the context of MORE-ADAPT-VQE.
As reference \cite{hong_refining_2024} points out, the trace of $\bar{H}$ is invariant to rotations among the references $\left\{\ket{\phi_i}\right\}$, meaning that for equal weights, minimizing the average expectation value of $\bar{H}$ in the basis of the references,
\begin{equation} \sum_i \omega_i E_i,\end{equation}
	is equivalent to minimizing the same average expectation value in the basis of the Ritz eigenstates,
\begin{equation} \sum_i \omega_i E_i'.\end{equation}
This is not true for the case of unequal weights.
This discrepancy highlights a choice we have made in defining MORE-ADAPT-VQE to minimize the former quantity rather than the latter.
In minimizing the former quantity, we would be pursuing the ``wrong'' objective function to some extent, though this is much cheaper than computing roughly $k^2$ off-diagonal gradient elements for each operator in each iteration, necessary for the latter possibility.
We explore the problems of weight selection and cost function definition in a manuscript currently in preparation.

Other ongoing work includes the calculation of more properties, exploration of reference selection strategies, and generalization of the method to approximate thermal ensembles.
Given these possible extensions of the method and its promising results in describing electronic excitations of small molecules, we believe that state-averaged ADAPT-VQE strategies warrants further exploration, both for applications in photochemistry and for developing a theoretical understanding of the limitations of other adaptive, excited-state VQEs.

\section{Data availability statement}
All data are available upon reasonable request.

\section{Code availability}
Code used in this project is available in the \textsc{QForte} package at:
\url{https://github.com/evangelistalab/qforte}

\section{Acknowledgements}
H.R.G. is grateful to Ayush Asthana for helpful discussions on q-sc-EOM and to Renke Huang, Ilias Magoulis, Jonathon Misiewicz, and Muhan Zhang for general consultation.
This work is supported by the U.S.\ Department of Energy under Award No.\ DE-SC0019374.

\section{Competing interests}
The authors declare no competing financial or non-financial interests.

\section{Author contributions}
Both authors contributed to the development of the theory.
 MC-VQE, MORE-ADAPT-VQE, and q-sc-EOM codes were implemented by H.R.G.
Data were obtained by H.R.G.
Both authors contributed to manuscript preparation.

\bibliographystyle{apsrev4-2}
\bibliography{untagged_revised_manuscript}
\end{document}